\documentclass[english]{revtex4}
\usepackage[T1]{fontenc}
\usepackage[latin9]{inputenc}
\usepackage{amsmath}
\usepackage{amssymb}
\usepackage{graphicx}
\usepackage{esint}

\makeatletter

\providecommand{\tabularnewline}{\\}

\@ifundefined{textcolor}{}
{%
 \definecolor{BLACK}{gray}{0}
 \definecolor{WHITE}{gray}{1}
 \definecolor{RED}{rg.b}{1,0,0}
 \definecolor{GREEN}{rgb}{0,1,0}
 \definecolor{BLUE}{rgb}{0,0,1}
 \definecolor{CYAN}{cmyk}{1,0,0,0}
 \definecolor{MAGENTA}{cmyk}{0,1,0,0}
 \definecolor{YELLOW}{cmyk}{0,0,1,0}
 }

\@ifundefined{showcaptionsetup}{}{%
 \PassOptionsToPackage{caption=false}{subfig}}
\usepackage{subfig}
\makeatother

\usepackage{babel}
\begin{document}

\title{Improved Criterion for Sawtooth Trigger and Modelling}

\author{A Zocco$^{1,2,*}$, J W Connor$^{1,2,3}$, C G Gimblett$^{1}$ and R J Hastie$^{1,2}$}

\affiliation{$^{1}$Euratom/CCFE Fusion Association, Culham Science Centre, Abingdon,
Oxon, OX14 3DB, UK\\
$^{2}$Rudolf Peierls Centre for Theoretical Physics, 1 Keble Road,
Oxford, OX1 3NP, UK\\
 $^{3}$Imperial College of Science Technology and Medicine, London
SW7 2BZ, UK}
\email{a.zocco1@physics.ox.ac.uk}
\begin{abstract}
We discuss the role of neoclassical resistivity and local magnetic
shear in the triggering of the sawtooth in tokamaks. When collisional
detrapping of electrons is considered the value of the safety factor
on axis, $q(0,t)$, evolves on a new time scale, $\tau_{*}=\tau_{\eta}\nu_{*}/(8\sqrt{\epsilon})$,
where $\tau_{\eta}=4\pi a^{2}/[c^{2}\eta(0)]$ is the resistive diffusion
time, $\nu_{*}=\nu_{e}/(\epsilon^{3/2}\omega_{te})$ the electron
collision frequency normalised to the transit frequency and $\epsilon=a/R_{0}$
the tokamak inverse aspect ratio. Such evolution is characterised
by the formation of a structure of size $\delta_{*}\sim\nu_{*}^{2/3}a$
around the magnetic axis, which can drive rapid evolution of the magnetic
shear and decrease of $q(0,t)$. We investigate two
possible trigger mechanisms for a sawtooth collapse corresponding
to crossing the linear threshold for the $m=1,~n=1$ instability and
non-linear triggering of this mode by a core resonant mode near the
magnetic axis. The sawtooth period in each case is determined by the time for the resistive evolution of the $q$-profile to reach the relevant stability threshold; in the latter case it can be strongly affected by $\nu_*.$
\end{abstract}
\maketitle

\section{Introduction}

When the safety factor, $q$, falls below unity on axis, tokamaks experience a ubiquitous periodic oscillation in the plasma core in which core parameters exhibit a ''sawtooth-like'' waveform, with a relatively slow ramp-up of, for example, the electron temperature,  followed by a rapid collapse. Understanding the characteristics of these sawtooth oscillations is important for predicting the performance of ITER since they can degrade the core confinement, expel the alpha particles that heat the burning plasma and couple to other instabilities that can severely limit operation.  The length of the sawtooth period, which is terminated by the sawtooth collapse, plays a major part in determining the impact of the sawtooth on the tokamak performance. Following a sawtooth collapse the radial profiles of the various plasma parameters evolve on transport timescales until a rapidly growing instability is triggered, producing magnetic reconnection on a fast timescale. It is this evolution, on the transport time scale, that determines the sawtooth period. 

The approach explored in this paper is to consider possible criteria for instability and, since these are sensitive to the $q(r)$-profile, to use a model for the resistive diffusion of $q$ following a sawtooth crash to monitor when these instability boundaries are crossed and, therefore, when the next sawtooth crash might be triggered. The time for this to occur yields the sawtooth period.

The resistive evolution is based on  neoclassical resistivity and builds on ideas of Park and Monticello\cite{parkmonticello} who realised that the evolution of the $q$-profile in neoclassical theory leads to a rapid cusp-like drop of $q$ on axis due to the effect of trapped electrons, with $q$ rapidly falling to  $q \sim 0.8$  in a sample MHD simulation. A more accurate treatment includes the collisional correction at small $\nu_{*e}=\nu_e/\epsilon^{3/2}\omega_{te}$ (here $\nu_e$ is the electron collision frequency, $\omega_{te}=v_{the}/R_0q$ the electron transit frequency and $\epsilon=a/R_0$ the inverse aspect ratio).  This removes the trapped particle cusp behaviour very close to the axis so that $dq/dr$  becomes zero there, although the value of $q$ is still rapidly driven well below unity. \\
This resistive evolution calculation requires an initial configuration given by the post-crash $q$-profile.  At present there is no generally accepted model for this. On the one hand it is recognised that the original "full-reconnection" model proposed by Kadomtsev\cite{kad2} cannot be accurate since direct measurements of $q_0$ indicate that, in most tokamaks, its value never rises to unity at any stage during the sawtooth cycle. On the other hand it is evident, from both MSE and Faraday rotation diagnostics, that some reconnection does occur during the turbulent conditions of sawtooth collapse. Some models of the sawtooth, however, assume that very little reconnection takes place \cite{igochine3}. In the present paper, we will assume that some reconnection does occur and, crucially, that the localised neoclassical peaking of the current density in the vicinity of the magnetic axis is destroyed, resulting in smooth behaviour of $J(r)$ and $q(r)$ in the plasma core. For simplicity, we take a Kadomtsev reconnected state as the post-crash initial condition for the evolution of $q$ during the sawtooth ramp, but this is not crucial for its evolution near the axis.

With this background we consider two potential instability models.
The first is in the spirit of the model proposed by Porcelli, Boucher and Rosenbluth\cite{0741-3335-38-12-010} which, in particular, proposes a condition on the magnetic shear at the $q = 1$ surface for triggering the sawtooth collapse. This condition is loosely connected with diamagnetic stabilisation of the internal kink mode. Here, we establish the stable window in an operating diagram for the $m = n = 1$ drift-tearing mode and the resistive internal kink mode, defined in terms of the plasma beta, $\beta=8\pi p/B^2$, and the instability drive, represented by a quantity $\Delta^{\prime}$ which is inversely related to the potential energy, $\delta W$, for the internal kink mode. This diagram results from an earlier study\cite{connor-hastie-zocco} of the stability of these two modes based on a plasma model with semi-collisional electrons and ions whose Larmor orbit exceeds the semi-collisional layer width where reconnection occurs. The trajectory of the core plasma state in this diagram as the profiles of $q$ and plasma pressure, $p$, change during the sawtooth ramp can be monitored and the triggering of the sawtooth crash identified as the point at which the linear stability threshold is crossed. This theory also predicts that the crash occurs when the magnetic shear, defined as $rq^\prime/q$, at $q = 1$, reaches a particular value, but is much more precisely defined than in the, somewhat heuristic, model of Porcelli et al.\cite{0741-3335-38-12-010}. 
The second model for the sawtooth period is based on the conjecture that an instability occurs on-axis if $q_0$ falls to some critical value, say $q_0=0.75$, when an $m=3$, $n=4$ ideal or tearing mode may be destabilised for example. Such an unstable mode can couple toroidally to a mode resonant at $q=1$, causing a sawtooth crash.
In the first model we monitor the shear at $q=1$ to determine the sawtooth period while in the second we follow the evolution of $q$ on axis. In the latter case we derive scaling laws for the period and discuss the implications for ITER.
  
The motivation for the present paper is, therefore, two-fold. Firstly we wish to explore, within a qualitative transport model, the importance of neoclassical evolution of the safety factor, $q(r,t)$, during the quiescent ramp phase of the sawtooth. As noted above, attention was first drawn to this by Park and Monticello\cite{parkmonticello} in their global sawtooth simulation, but although this phenomenon should be present in the sawtooth modelling of Ref.\cite{0741-3335-38-12-010} and others, there has been little discussion of its significance in these studies.

The second purpose of this paper is to replace the heuristic stability boundaries ( i.e. the sawtooth trigger conditions) employed in Ref.\cite{0741-3335-38-12-010} by analytic marginal stability conditions derived in \cite{connor-hastie-zocco}. This has the effect of replacing unknown coefficients in Ref.\cite{0741-3335-38-12-010} by precise values with appropriate functional dependencies on parameters such as $\eta_e$, $\eta_i$, etc. We note here that  no attempt is made to calculate the tearing stability index, $\Delta^\prime$, or the closely related quantity $\delta W$, with its important dependence on contributions from energetic ion populations in the plasma core.  Such quantities are taken as given. In a thorough implementation of the present ideas these quantities would need to be evaluated in a separate calculation and a full $1\frac{1}{2}-D$ transport code, as in \cite{0741-3335-38-12-010}, would be required. Of course $\delta W$ will also evolve during the long quiescent ramp phase of a sawtooth but, in what follows, we assume that $\delta W$ may have reached a quasi steady state and that the resistive evolution of $q_0$, or, of $r_1 q^\prime(r_1)$, may have a crucial influence in triggering the next sawtooth collapse.  Evidence in support of this picture appeared in the, very effective, triggering of a sawtooth collapse using ECCD in ASDEX\cite{igochine,manini}. 

\section{Evolution of the $q(r,t)$-profile during the Sawtooth ramp}
\subsection{Resistive evolution model}
In this Section, we study the importance of the trapped particle correction to Spitzer resistivity in determining the duration
of the sawtooth ramp. During this period, which follows a collapse
event, thermal equilibrium can be assumed to be rapidly re-established, but the current
profile, and the $q$-profile, evolve resistively towards a remote
(and ideal MHD unstable) steady-state with $q_0<1/2$ \cite{BussacdeltaW} in which the toroidal current is $J_{\phi}=E_{0}/\eta(r),$
with $E_{0}$ constant, and $\eta(r)$ the resistivity. For the moment
we ignore the effect of the Bootstrap current within Ohm's law.
Neoclassical resistivity is given approximately by \cite{hirshmann,sauter}
\begin{equation}
\eta(r)=\eta_{Sp}(r)/(1-\sqrt{r/R_{0}})^{2},\label{eq:resneo}
\end{equation}
where $\eta_{Sp}$ is the Spitzer resistivity. Assuming the electron temperature
profile to be given by $T_{e}(r)=T_{0}(1-r^{2}/a^{2})^{4/3},$ the Spitzer
resistivity has the form 
\begin{equation}
\eta_{Sp}(r)=\frac{\eta_{0}}{\left(1-r^{2}/a^{2}\right)^{2}}.
\end{equation}
We construct the relevant diffusion equation for the $q$-profile
in the cylindrical tokamak limit retaining one toroidal effect, namely
the neoclassical correction to resistivity. Thus, 
\begin{equation}
\begin{split}\frac{\partial B_{\theta}}{\partial t} & =-c\left(\nabla\times\mathbf{E}\right)_{\theta}=c\frac{\partial}{\partial r}\left(\eta J_{z}\right)=\frac{\partial}{\partial r}\left[\frac{\eta c^{2}}{4\pi r}\frac{\partial}{\partial r}\left(rB_{\theta}\right)\right],\end{split}
\label{eq:bpoloidal}
\end{equation}
 and using the definition of the safety factor, 
\begin{equation}
q(r)=\frac{r}{R_{0}}\frac{B_{z}}{B_{\theta}},
\end{equation}
 this becomes 
\begin{equation}
\frac{\partial}{\partial \tau}\left(\frac{1}{q}\right)=\frac{1}{r}\frac{\partial}{\partial r}\left[\frac{\hat\eta}{r}\frac{\partial}{\partial r}\left(\frac{r^{2}}{q}\right)\right],\label{eq:qnotnorm}
\end{equation}
where we have introduced the dimensionless variables defined by $\tau=t/\tau_{\eta},$
and $r=r/a,$ with $\tau_{\eta}=4\pi a^{2}/(\eta_0 c^{2}).$ The model
for neoclassical resistivity is thus
\begin{equation}
\hat{\eta}(r)=\frac{1}{\left[(1-r^{2})(1-\sqrt{\epsilon}r^{1/2})\right]^{2}},
\end{equation}
where $\epsilon=a/R_{0}.$ Clearly, the fractional power in the trapped
electron correction to Spitzer resistivity generates (unphysical)
singular behaviour in Eq. \eqref{eq:qnotnorm}, for $r\rightarrow0,$
i.e. in the vicinity of the magnetic axis. This is removed by including
the transition from a neoclassical resistivity to Spitzer when 
\begin{equation}
\nu_{e}>\frac{v_{the}}{R_{0}q}\left(\frac{r}{R_{0}}\right)^{3/2}.
\end{equation}
 Incorporating this correction, the expression for the resistivity
becomes 
\begin{equation}
\hat{\eta}(r)=\frac{1}{\left[(1-r^{2})(1-\frac{\sqrt{\epsilon}r^{2}}{r^{3/2}+\nu_{*}})\right]^{2}},\label{eq:resnorm}
\end{equation}
where $\nu_{*}=\nu_{e}/(\epsilon^{3/2}\omega_{te}),$ with $\omega_{te}=v_{the}/(R_0 q).$  
In large tokamaks such
as JET or ITER, the dimensionless parameter $\nu_{*}$ is extremely
small, so that resistive evolution in the vicinity of the magnetic
axis, though not singular there, is likely to be rapid: this will become
evident from our numerical solution of Eq. \eqref{eq:qnotnorm}. Furthermore,
although the scaling of the resistive diffusion time, $\tau_{\eta}\propto a^{2}T_{e}^{3/2}$
points to a much slower evolution of $q(r,t)$ in ITER than in JET (possibly
by a factor of $\sim60$), the scaling of the small parameter, $\nu_{*}\propto N_{e}a/T_{e}^{2}$
reduces this factor when considering core evolution times. For example,
by expanding Eq. \eqref{eq:qnotnorm} locally around $x=0,$ and employing
Eq. \eqref{eq:resnorm}, one obtains the solution 
\begin{equation}
q_{0}(t)=q_{0}(0)\exp\left(-t/\tau_{*}\right),
\end{equation}
 with 
\begin{equation}
\tau_{*}=\tau_{\eta}\frac{\nu_{*}}{8\sqrt{\epsilon}}\propto\frac{R_{0}^{3}N_{e}}{T_{e}^{1/2}}.\label{eq:taustar}
\end{equation}
 Hence, at early times, the safety factor undergoes an exponential
decay on the timescale $\tau_{*}.$ Note that the presence of the
short timescale $\tau_{*}$ is the consequence of the formation in
the $q$-profile of a boundary layer of width $\delta_{*}\sim\nu_{*}^{2/3}a,$
which we assume to be destroyed by the crash itself, and thus, not 
present at $t=0$; it develops only afterwards.

Since Eq. \eqref{eq:qnotnorm} is of the heat diffusion type, in order
to solve it we need an initial value over the whole domain $r\in[0,1],$
and two boundary conditions at $r=0$ and $r=1.$ 
We then choose the Cauchy boundary condition $q(1,t)=q_{in}(1),$
i.e. the total plasma current is held constant, and the Neumann
boundary condition $\partial_{r}q^{-1}(0,t)=0.$ The last one is chosen
because we want our system to evolve, at very long times, towards
an equilibrium magnetic field which is regular for $r\rightarrow0.$
To clarify this point, let us consider the case of constant resistivity.
After setting $\partial_{t}\equiv0$ and integrating Eq. \eqref{eq:qnotnorm} twice,
one obtains the equilibrium safety factor
\begin{equation}
q_{eq}^{(0)}(r)=\frac{r^{2}}{C_{2}+C_{1}r^{2}},\label{eq:qeqconsteta}
\end{equation}
where $C_{1,2}$ are two constants of integration. Then, if $C_{2}\neq0$
for $r\rightarrow0,$ we have $q_{eq}(r)\rightarrow C_{2}^{-1}r^{2},$
which implies a divergent magnetic field, $B_{\theta}/B_{0}\equiv B\sim C_{2}/r$
for $r\rightarrow0$ . Hence, we set $C_{2}\equiv0$. Then $q_{eq}(r)\rightarrow C_{1}^{-1}$
for $r\rightarrow0,$ which yields $B/r\sim C_{1}$ for $r\rightarrow0.$
From this it follows that $\partial_{r}(B/r)=0,$ or equivalently $\partial_{r}q^{-1}(0,t)=0.$
The result in Eq. \eqref{eq:qeqconsteta} can be easily generalised
to the case of non-constant resistivity, giving 
\begin{equation}
q_{eq}(r)=q_{pc}(r=1)\left(\int_{0}^{1}\frac{\varrho d\varrho}{\hat{\eta}(\varrho)}\right)\frac{1}{\int_{0}^{r}\frac{\varrho d\varrho}{\hat{\eta}(\varrho)}}.\label{eq:qeqgen}
\end{equation}
The integrals in Eq. \eqref{eq:qeqgen} can be performed analytically and then
the result can be compared to the long time evolution given by Eq. \eqref{eq:qnotnorm}.
Figure $2(a)$ shows the solution of Eq.5 starting from an arbitrarily chosen initial state with $q(r)=1$ everywhere,  for $t/\tau_{*}=120,\,200,\,300$. Also shown is
 the analytical solution $q_{eq}(r)$ for $\nu_{*}=10^{-4}:$ the
analytical steady-state equilibrium is recovered. It is worth noticing that
it is reached on a time which is much shorter than the resistive diffusion
time, $\tau_{\eta}=5\times10^{4}\tau_{*}.$ 
\begin{figure}
\subfloat[The numerical solution of Eq. \eqref{eq:qnotnorm} for $t/\tau_{*}=120,\,200,\,300.$
The analytical solution $q_{eq}(r)$ is approached from above as time
increases. Here $\nu_{*}=10^{-4}.$]{

\includegraphics[scale=0.5]{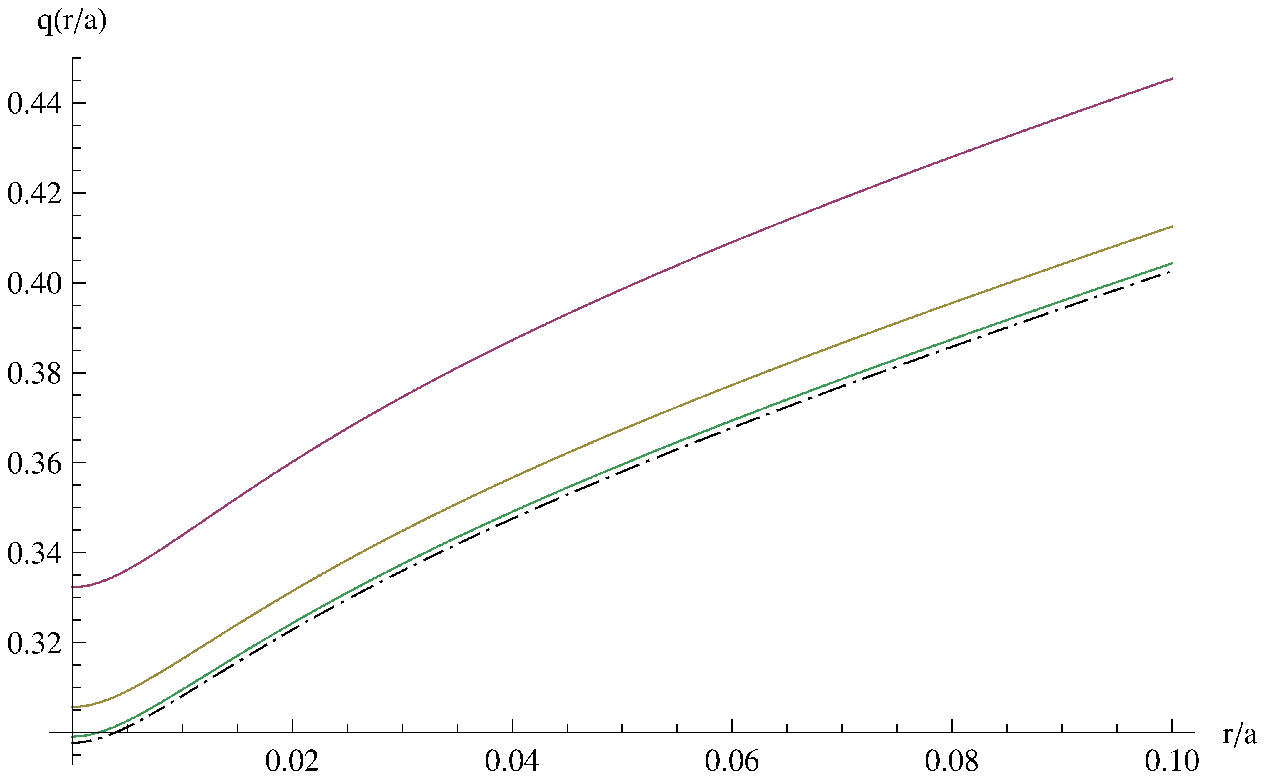}\label{fig:kadeq}}\subfloat[The electric field $E=\hat{\eta}r^{-1}\partial_{r}(r^{2}/q)$ calculated
from the numerical solution of Eq. \eqref{eq:qnotnorm} for $t/\tau_{*}=120,\,200,\,300.$
The constant solution is approached as time increases. Here $\nu_{*}=10^{-4}.$]{\includegraphics[scale=0.5]{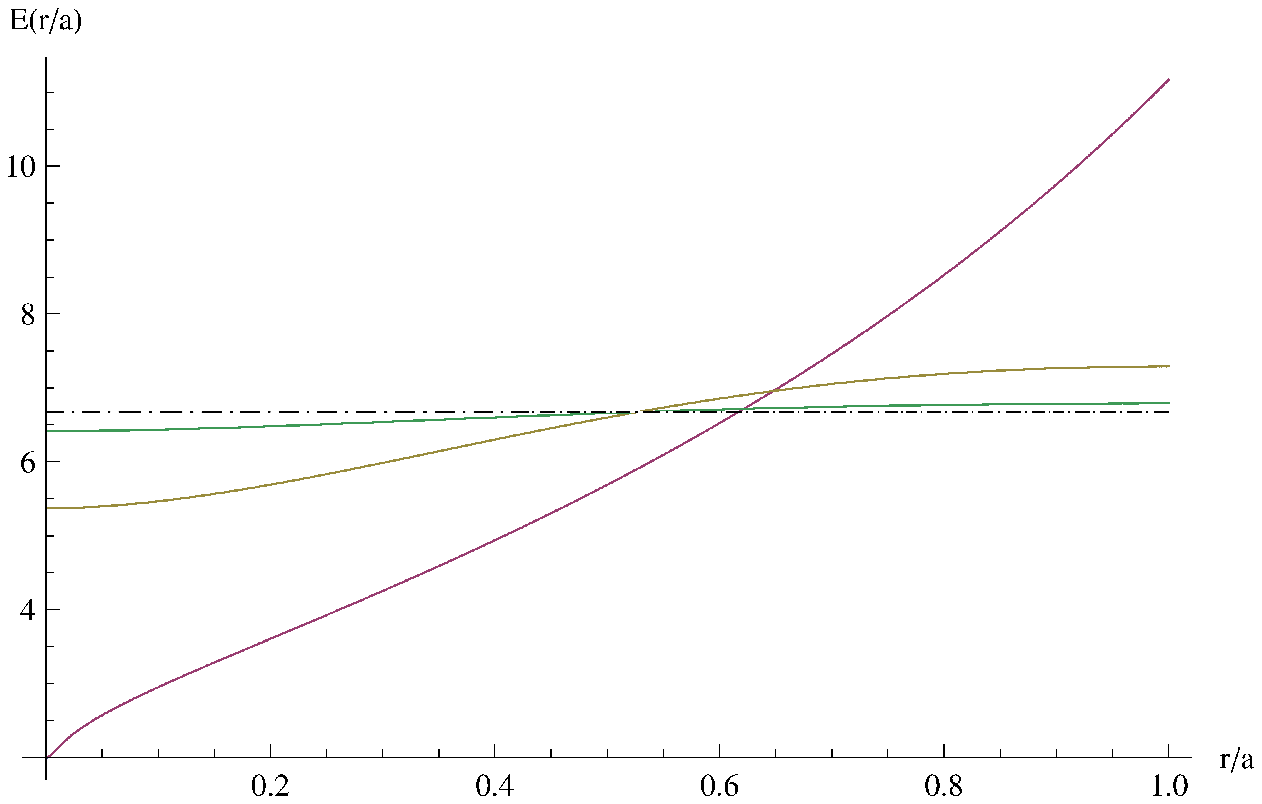}\label{fig:Eleq}

}

\label{fig:fig3-1}
 \caption{Long time evolution of Eq. \eqref{eq:qnotnorm} for $q$-profile and electric field.}
\end{figure}
Another important property of the steady-state solution of Eq. \eqref{eq:qnotnorm}
is that it must give a (radially) constant electric field {[}see Eq.
\eqref{eq:bpoloidal}{]}. In Fig. \ref{fig:Eleq} we show the electric
field $E=\hat{\eta}r^{-1}\partial_{r}(r^{2}/q)$ calculated from the
solutions in Fig. \ref{fig:kadeq}. As expected, the final state is
uniform through the domain of integration.

\subsection{Evolution from a fully-reconnected Kadomtsev-like state}

As an initial $q(r)$ state we first choose, for simplicity, the fully reconnected Kadomtsev-like
form, $q_{pc}(r)$ given by: 
\footnotetext[1]{The $\tanh$ functions have only been inserted to provide a slight
spread of the initial current sheet of width $\varsigma$ at $r_{mix}$
in the Kadomtsev model. The choice of the pre-crash is $q(r)=q_0/[1-r^2+(1/3)r^{4}].$ This is actually the steady-state $q$ if resistivity were Spitzer
and $T_{e}\propto(1-r^{2})^{4/3}$. } 
\begin{equation}
\begin{split} & \frac{2}{q_{pc}}=\frac{1-\tanh\left[(r-r_{mix})/\varsigma\right]}{q_{K}}+\frac{1+\tanh\left[(r-r_{mix})/\varsigma\right]}{q_{in}},\end{split}
\label{eq:kad_last}
\end{equation}
 where $q_{K}=1/(1-0.27r^2)$ is the Kadomtsev fully-reconnected state which has been calculated numerically for initial values $q_0=0.75$ and $q_{in}(r)=q_{0}/[1-r^2+(1/3)r^{4}]$ chosen for the ''pre-collapse'' state, resulting in
$r_{mix}=\sqrt{9-\sqrt{144q_{0}-63}}/2\approx0.757$.  $\varsigma=5\times10^{-3}$ represents the narrow width of the current sheet at the mixing radius, $r_{mix}.$ The profiles $q_{in}(r)$ and $q_{pc}(r)$ are  shown in Fig.1
\begin{figure}
\includegraphics[scale=0.8]{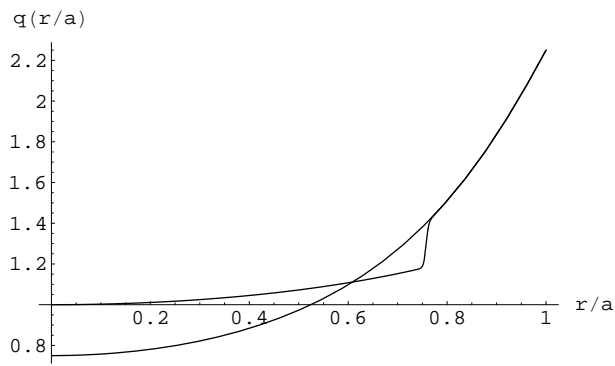} \caption{Kadomtsev-like pre-crash $q_{in}=q_{0}/[1-r^2/a^2+(1/3)r^{4}/a^4]$ and post-crash
$q_{pc}$-profile as given by Eq. \eqref{eq:kad_last}.}
\label{fig:fig2} 
\end{figure}

\begin{figure}
\includegraphics[scale=0.8]{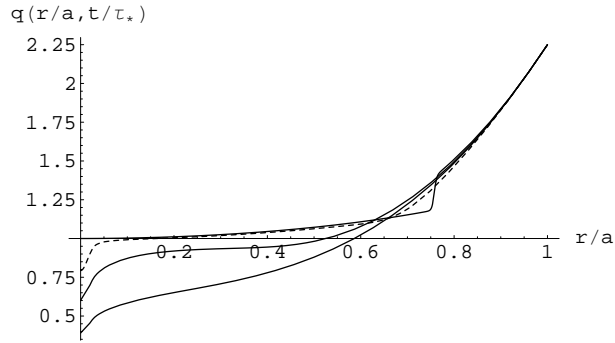} \caption{Solution of Eq. \eqref{eq:qnotnorm} for time $t/\tau_*=0,$ $2.5\,,$
$37.5\,,$ and $49.75\,.$ Here $\nu_{*}=10^{-3}.$}
\label{fig:fig3-2} 
\end{figure}
In Fig. \ref{fig:fig3-2} we show an example of the evolution
of the whole $q$-profile, starting from the Kadomtsev post-crash
state. The formation of the structure on the width $\delta_{*}$ is evident. 
It is also clear that a 
very fast diffusion of the initial current sheet occurs
at the mixing radius $r_{mix}.$ 
In tokamaks, such a fast neoclassical evolution of the $q$-profile seems to occur only when approaching stationary conditions \cite{lloyd, kelliher}.  
Here we also note that the final equilibrium of Eq. \eqref{eq:qeqgen} is ideal MHD unstable
to the $m=1,$ $n=2$ mode (since $q_0<1/2$ \cite{BussacdeltaW}),
so from an operational point of view, it should never be achieved!

Our results regarding the evolution of $q$ in the axial region are not sensitive to the assumption of a Kadomtsev-like post crash state; a Taylor relaxation model of Ref. \cite{chris_gimblett3} would lead to similar results.  The case with $q(r)=1$ everywhere as an initial condition, shown in Fig.2, provides another example. The key point is that there is sufficient reconnection near the axis to interrupt the cusp-like neoclassical evolution of $q$. If this were absent and the only changes to $q$ arose as a consequence of sawtooth oscillations in temperature affecting the resistivity profile, the axial value of $q$  would ultimately saturate at some low value, $q_0<0.5$, due to the neoclassical evolution, contrary to observation. On the other hand, the Kadomtsev model is somewhat special in that the $q=1$ surface appears at $r=0$ after the crash, whereas in the Taylor relaxation model it is much further out; consequently the evolution of the shear at $q = 1$ is rather different in the two cases.

\section{Linear Stability and axial criterion for the Sawtooth Trigger}

\subsection{The role of the drift-tearing and kink modes }

In Ref. \cite{connor-hastie-zocco} we developed a unified theory
of the drift-tearing mode and internal kink mode relevant to the $m=1,~n=1$
mode resonant at $q=1$ in large hot tokamaks such as ITER. Specifically,
we adopted a plasma model with semi-collisional electrons for which:
\begin{equation}
k_{\parallel}^{2}v_{the}^{2}\sim\omega\nu_{e}
\end{equation}
 with $k_{\parallel}\equiv k_{y}x/L_{s}$, the parallel wavenumber, $k_{y}$ the component
of the perpendicular wavenumber lying within the magnetic surface,
$x$ the distance from the resonant surface, $L_{s}=R_0q/s$, the
shear length, and $\omega$ the mode frequency. The resulting width of the electron current channel
$\delta$ is thus given by: 
\begin{equation}
\delta=\frac{(\omega\nu_{e})^{1/2}L_{s}}{k_{y}v_{the}}.
\end{equation}
 We consider the ion Larmor orbit to be large, $\rho_{i}\gg\delta.$
The theory is characterised by two other key parameters : $\hat{\beta}=0.5\beta_{e}L_{s}^{2}/L_{n}^{2}$
and $r_1\Delta^{\prime}=\hat{s}_{1}^{2}/\delta W$, where $L_{n}^{-1}=-N_{e}^{-1}\partial_{r}N_{e}$
is the inverse of the equilibrium electron density gradient length, and $\Delta^{\prime}$
is the instability drive \cite{FKR}. For the $m=1,~n=1$ modes, this
is related to the potential energy of the internal kink mode, $\delta W$.
The key results are that, at low values of $\hat{\beta}$, the drift-tearing
mode, with frequency $\omega=\omega_{*e}(1+0.73\eta_{e})$, is stabilised
by finite ion orbit and diamagnetic effects, provided that {[}see
Eq. (42) of \cite{connor-hastie-zocco}{]}: 
\begin{equation}
\Delta^{\prime}<\Delta_{1}^{\prime},
\end{equation}
 where 
\begin{equation}
\rho_{i}\Delta_{1}^{\prime}=\sqrt{\pi}\hat{\beta}\frac{(\hat{\omega}-1)^{2}(\hat{\omega}\tau+1)(\hat{\omega}\tau+1-\eta_{i}/2)}{\hat{\omega}^{2}(1+\tau)^{2}}log(\Lambda)-\pi\hat{\beta}(\hat{\omega}-1)\bar{I},
\end{equation}
 with 
\begin{eqnarray}
\Lambda=e^{-\frac{\pi}{4}}\frac{\rho_{i}}{\delta_{0}\,\hat{\omega}^{1/2}},\\
\hat{\omega}=1+0.73\eta_{e}.
\end{eqnarray}
 Here $\hat{\omega}=\omega/\omega_{*e},$ $\omega_{*e}=1/2k_{y}v_{the}\rho_{e}/L_{n},$
$\bar{I}$ is an integral defined in Ref. \cite{connor-hastie-zocco} with
approximate value $\sim-\eta_{e}^{1/2}$, $\tau=T_{e}/T_{i},$ $\delta_{0}=\sqrt{\omega_{*e}\nu_{e}}r_{1}L_{s}/v_{the},$
and $r_{1}$ is the position at which $q=1.$ This result is valid
at small $\Delta^{\prime}\rho_{i}\sim\hat{\beta}.$ At higher values
of $\Delta^{\prime}\rho_{i}$ , i.e. as $\hat{\beta}\Delta^{\prime}\rho_{i}\sim1$
the unstable drift-tearing mode couples to a stable Kinetic Alfven
Wave (KAW) until, at a critical value of the parameter $\hat{\beta}\Delta^{\prime}\rho_{i},$
there is an exchange of stability, with the drift-tearing mode continuing
at the same frequency but now stable, whereas the (previously stable)
KAW becomes unstable. With continuing increase of $\hat{\beta}\Delta^{\prime}\rho_{i}$
the KAW frequency drops towards $\omega/\omega_{*e}=1$ and its growth
rate also decreases, the mode eventually becoming stable when {[}see
Eq. (49) of \cite{connor-hastie-zocco}{]} 
\begin{equation}
\Delta^{\prime}>\Delta_{2}^{\prime},
\end{equation}
 with 
\begin{equation}
\rho_{i}\Delta_{2}^{\prime}=2.42\pi\frac{\rho_{i}}{\delta_{0}}\frac{\eta_{e}\hat{\beta}}{\sqrt{5.08+2\sqrt{2.13}-1.71\eta_{e}/(1+\tau)}}.
\end{equation}
These results pertain to low $\hat{\beta},$ however, it has also
been shown that the KAW is stable when $\hat{\beta}$ exceeds a critical
value depending on $\eta_{e}$, due to the effect of shielding of
the resonant surface by plasma gradients. For the particular case,
$\eta_{e}=2.53$, this threshold is $\hat{\beta}=0.34$. Such a result
is consistent with that of Drake et al.\cite{drake:2509} who found
stability in the limit $\hat{\beta}\sim(\hat{\omega}-1)^{-1}\gg1,$ for $\tau\gg 1$

In Ref. \cite{connor-hastie-zocco}, the stability limit for the
dissipative internal kink mode was also determined. In particular,
for $\delta W>0,$ one stability limit is given by, 
\begin{equation}
\Delta^{\prime}<\Delta_{3}^{\prime},
\end{equation}
 where 
\begin{equation}
\Delta_{3}^{\prime}\delta_{0}=\hat{\beta}\frac{\pi^{2}}{\sqrt{\frac{1+\tau}{4.26}(4.08-1.71\eta_{e})}}\frac{1}{\log\left[\hat{\beta}^{2}\frac{\rho_{i}}{\delta_{0}}\frac{\pi}{(1+\tau)^{3/2}}\sqrt{\frac{4.26}{4.08-1.71\eta_{e}}}\right]+\pi-0.5}\label{eq:delprime}
\end{equation}
 {[}see Eq. (96) of Ref. \cite{connor-hastie-zocco}, here we are
taking $\eta_{i}=0$, hence $I_{2}\approx-0.5${]}, while a general
stability boundary for arbitrary $\hat{\beta}$, which ensured stability
when $\hat{\beta}>\hat{\beta}_{2}\approx\sqrt{\delta_{0}/\rho_{i}},$
was also determined.

\begin{figure}
\includegraphics[scale=0.7]{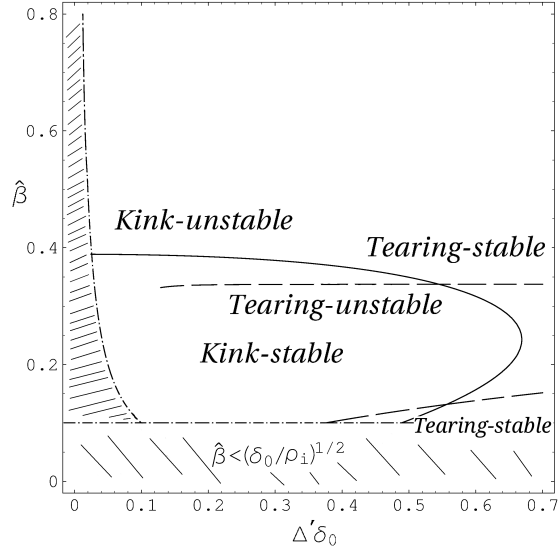}
\caption{Stability boundary for the drift-tearing mode and kink mode as derived
in Ref. \cite{connor-hastie-zocco} in the space of $\hat\beta$ and $\Delta^{\prime}\delta_0$. Here $\delta_{0}/\rho_{i}=0.01.$ Solid and dashed lines are the boundaries for the kink and tearing modes, respectively. The theory requires $\delta_0/\rho_i<\hat\beta^2$ and $\hat \beta \gtrsim 1/(\Delta^{\prime}\rho_i).$ For $(\delta_0/\rho_i^{1/2})<\hat\beta,$ the two boundaries cross at $\delta_0 \Delta^{\prime}\sim 0.5.$ For $\delta_0 \Delta^{\prime}\lesssim 0.5,$ we have a stable region [second inequality in Eq. ($24$)].}
\label{fig:geometry-1-1} 
\end{figure}

The combined effects of these stability boundaries is encapsulated
in Fig. \ref{fig:geometry-1-1}. In particular, when $\hat{\beta}\ll1$,
we observe the existence of two stable ranges for the stability index
$\Delta^{\prime}$; namely 
\begin{equation}
\Delta^{\prime}<\Delta_{1}^{\prime},\,\,\,\,\,\,\,\,\Delta_{2}^{\prime}<\Delta^{\prime}<\Delta_{3}^{\prime}.
\end{equation}
 Thus, as Fig. \ref{fig:geometry-1-1} clearly shows, we can identify
a new window in \textbf{$\hat{\beta}$} for the instability for the
drift-tearing mode that was absent from previous calculations that exploited
arbitrarily large\textbf{ $\hat{\beta}$} \cite{drake:2509}.
It is crucial to understand the relative magnitude of $\Delta_{2}^{\prime}$
and $\Delta_{3}^{\prime}.$ In particular, for $\Delta_{2}^{\prime}/\Delta_{3}^{\prime}<1,$
the low-$\hat{\beta}$ window of stability is accessible to the system.
If we define$f(\eta_{e})=\Delta_{2}^{\prime}/\Delta_{3}^{\prime}-1,$
we can solve for $f(\eta_{e})=0,$ finding that the window of stability
is present when $f(\eta_{e})<0.$
\begin{figure}
\includegraphics[scale=0.7]{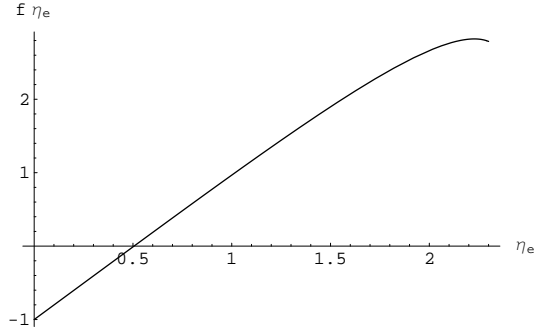} \caption{Function $f=f(\eta_{e})$ for $\hat{\beta}^{2}\rho_{i}/\delta_{0}=10,$ and $\tau=1$ }
\label{fig:window-1} 
\end{figure}
In Fig. \ref{fig:window-1} we show the plot of $f(\eta_{e});$ we
see that there is a critical electron temperature gradient, $\eta_e^{(0)},$ for which
\begin{equation}
\Delta_{2}^{\prime}/\Delta_{3}^{\prime}\lessgtr1,\,\,\,\mbox{\mbox{when}}\,\,\eta_{e}\lessgtr\eta_{e}^{(0)}.
\end{equation}
 Some values of the new critical electron gradient $\eta_{e}^{(0)}$
are given in Table \ref{tab:tab1}. The window of instability only exists for a narrow range of values of $\eta_e:  \eta_e\lesssim 0.4.$
\begin{table}
\begin{centering}
\begin{tabular}{|c|c|c|c|c|}
\hline 
$\eta_{e}^{(0)}$  & $0.52$  & $0.50$  & $0.46$  & $0.42$\tabularnewline
\hline 
$\hat{\beta}^{2}\rho_{i}/\delta_{0}$  & $9$  & $10$  & $15$  & $25$\tabularnewline
\hline 
\end{tabular}
\par\end{centering}
\caption{Critical electron temperature gradient $\eta_{e}^{(0)}$ as a function
of $\hat{\beta}^{2}\rho_{i}/\delta_{0}$, given by the solution of
the equation $f(\eta_{e})=0.$}
\label{tab:tab1} 
\end{table}

Once all the marginal stability boundaries for these modes have been determined,
it remains to understand how they can be crossed. 
The shear dependence of the key
parameters $\hat{\beta}$ and $\delta$, reveals great sensitivity
of the various thresholds to an evolving $q(r)$ profile. In particular,
assuming a parabolic density profile $N_e=N_{0}(1-r^{2}/a^{2})$ and
$T_{i}=T_{e},$ [see Eq. (109) of Ref. \cite{connor-hastie-zocco}],
for a given inverse aspect ratio $\epsilon,$ we have 
\begin{equation}\label{eq:critbeta}
\hat{\beta}=\hat\beta_c=\frac{\beta_{0}}{\epsilon^2\,{[aq^{\prime}(r_{1})]}^{2}}\approx0.5/{[aq^{\prime}(r_{1})]}^{2},
\end{equation}
 in JET or ITER. 
Bearing this is mind we find $\Delta^{\prime}_1\propto \hat s_1^4,$ while $\Delta^{\prime}_2$ and $\Delta^{\prime}_3$ are both proportional to $\hat s_1^3.$ Furthermore, Eq. (\ref{eq:critbeta}) shows that the screening threshold, $\hat \beta\approx \hat \beta_c,$ is also sensitive to $\hat s_1^2.$

From this analysis, a new picture of the boundaries of linear marginal
stability, and their use for sawtooth modelling, emerges. This differs from that of Ref. \cite{0741-3335-38-12-010}. While the stability criteria associated to $\Delta^{\prime}_1$ and $\Delta^{\prime}_2$ do not appear in Ref. \cite{0741-3335-38-12-010}, the $\Delta^{\prime}_3$ threshold is equivalent to Eq. $(15)$ of Ref. \cite{0741-3335-38-12-010}:
\begin{equation}\label{eq:oldcrit}
-c_{\rho}\rho_{i}/r_{1}<-\delta W,\,\,\,\,\mbox{and}\,\,\,\omega_{*i}<c_*\gamma_K.
\end{equation}
These were introduced as heuristic conditions for a generally large $\Delta^{\prime},$ where we note the relationship
\cite{ABC} 
\begin{equation}
\Delta^{\prime}r_{1}=\frac{\hat{s}_{1}^{2}}{\delta W},
\end{equation}
Here $c_{\rho}$ is a phenomenological constant,
$\omega_{*i}$ the ion diamagnetic frequency, $\gamma_{K}$ the
growth rate of the dissipative kink mode, $c_{*}$ another phenomenological
constant, and $\hat{s}_{1}=r_1,$ $q^{\prime}\left(r_1\right)$ is the magnetic shear at the $q=1$ surface. The energy integral
$\delta W$ can include energetic particle contributions, but in their absence, and in the large aspect ratio tokamak limit, it is given by $\delta W=(r_{1}^{2}/R_{0}^{2})\delta\tilde{W}_{T},$ with $\delta\tilde{W}_{T}$
the energy calculated by Bussac et al. \cite{BussacdeltaW}. However, 
whereas the $\Delta^{\prime}_3$ criterion is appropriate when $\delta W >0,$ condition (\ref{eq:oldcrit}) may require $\delta W<0, $ i.e. an unstable ideal mode.
To relate this model to our results, we rewrite Eq. \eqref{eq:delprime} in the following way
\begin{equation}\label{eq:porcodio}
-\hat{s}_{1}^{3}\frac{a}{\rho_{i}}\frac{a}{R_0}\frac{a}{r_1}\sqrt{\frac{0.5\nu_{e}}{\Omega_{e}}}\frac{\sqrt{\frac{1+\tau}{4.26}(4.08-1.71\eta_{e})}}{2\pi^{2}\beta_{e}}\frac{\rho_{i}}{r_{1}}<-\delta W.
\end{equation}

From Eq. (\ref{eq:porcodio}) we obtain two important results. The first
is that there is an exact relationship between the critical shear for instability
and the ideal MHD potential energy: 
\begin{equation}
\hat{s}_{crit}^{3}\approx\delta W\frac{R_{0}}{a}\frac{r_{1}^2}{a^2}\sqrt{\frac{\Omega_{e}}{0.5\nu_{e}}}2\pi^{2}\frac{\beta_{e}}{\sqrt{\frac{1+\tau}{4.26}(4.08-1.71\eta_{e})}}.\label{eq:critshearCHZ}
\end{equation}
Thus the critical shear is not a simple constant, but scales like
$\hat{s}_{crit}\propto\delta W^{1/3}.$ The second is the electron
temperature gradient dependence of such threshold. This aspect was already
considered in the literature \cite{sauter2,angioni}, but not
derived analytically.

\subsection{Axial criterion}

One alternative model to trigger a sawtooth might be related to the axial evolution of the safety factor $q$. In fact, $q_{0}$ can undergo a rapid downward evolution during the ramp that precedes the crash. Thus, one might consider what can limit
such evolution of $q$ on axis. It is known that, even before the ideal MHD $m=1,$ $n=2,$
instability becomes possible (when $q_0<1/2$), the tearing mode stability index $\Delta^{\prime}_{m,n}$ 
for core resonant modes, such as $m=2,$ $n=3,$ or $m=3,$ $n=4,$ can become positive and potentially unstable, see for example Fig $6.9.2$ of Ref. \cite{wessonbook}. Furthermore, diamagnetic stabilisation is likely to be extremely weak close to the magnetic axis, and the average curvature is unfavourable \cite{10.1063/1.861224}. Hence, it is tempting to look
for a correlation between the onset of such modes, and the sawtooth
period. 

\section{Sawtooth Period}

\subsection{Time evolution of shear at the $q=1$ rational surface and $q_0$}

In Section 3, we stressed the fact that both $\hat{\beta}$ and $\Delta^{\prime},$
which are fundamental to the calculation of the boundary of marginal
stability of the drift-tearing and kink modes, show a shear dependence.
Hence, we present in Figs.6 and 7 the time evolution of:  $q_0,$ the value of $q$ on
axis, $r_1,$ the position of the $q(r_{1},t)=1$ surface,  the shear at the $q=1$ surface, $\hat{s}_{1}=r_{1}q^{\prime}(r_{1}),$ and finally the parameter $\hat{\beta}=0.5/[aq^{\prime}(r_{1})]^{2}.$ 

\begin{figure}[!h]
\subfloat[]{\input{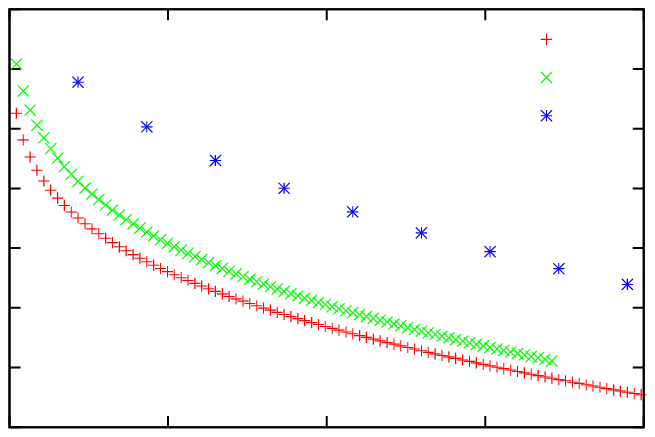}\label{fig:qnotplot}}
\subfloat[]{\input{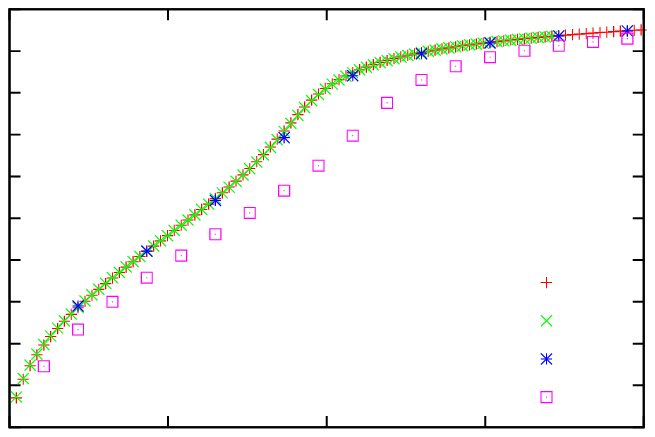}\label{fig:r1plot}}
\label{fig:fin}
 \caption{The value of $q$ on axis and the position of the $q=1$ surface calculated from Eq. \eqref{eq:qnotnorm} for different $\nu_*$.}
\end{figure}

\begin{figure}[!h]
\subfloat[]{\input{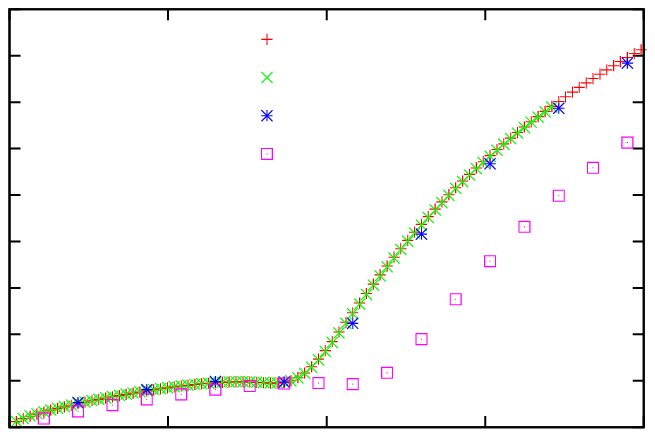}\label{fig:sheaplot}}
\subfloat[]{\input{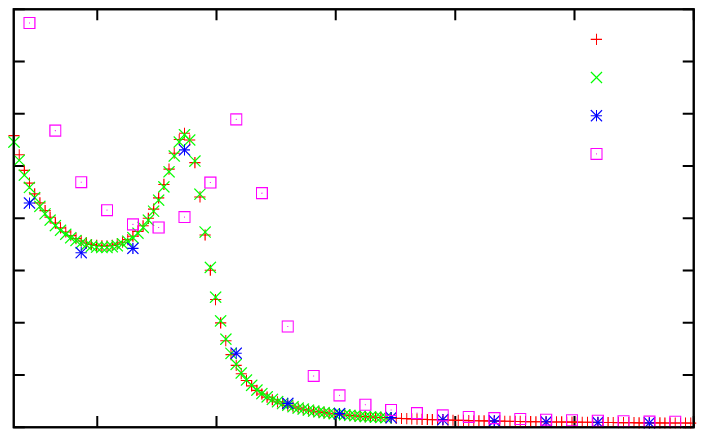}\label{fig:betaplot}}
\label{fig:fin1}
\caption{The shear at the resonant surface $\hat s_1$ and the parameter $\hat\beta$ calculated from Eq. \eqref{eq:qnotnorm} for different $\nu_*$.}
\end{figure}

Figures $6(b),$ and $7$  indicate that there is a negligible dependence of the evolution of $r_1(t),$ $\hat s_1(t),$ and $\hat{\beta}(t)$ on the collisionality parameter $\nu_{e*}$ in the range $\nu_* < 10^{-2}$. From transport modelling simulations, we know that a possible value
of critical shear at which the sawtooth is expected to be triggered
is$^1$ $\hat{s}_{1}=0.4,$ \cite{chapman3}
\footnotetext[1]{We note that from the literature one cannot refer to a typical
value. Many results are obtained by modelling activities and give
values that range from $0.15$ for TCV \cite{angioni} to up to $0.6$ \cite{chapman3} .}
However, we remind the reader that the critical shear found in this
work is not a simple function of plasma and machine parameters, but
scales with the one-third power of the MHD energy $\delta W,$ and this
quantity is inevitably expected to vary during the sawtooth ramp,
complicating our picture. On the other hand, it has been known that
a simple condition $\hat{s}_{1}=\mbox{const.}$ can fail to reproduce
the observed variations of the sawtooth period in response to localised
electron cyclotron heating and current drive \cite{angioni}. At
present, it remains an open question whether the failure of a condition
$\hat{s}_{1}=\hat{s}_{crit}\equiv\mbox{const.}$\cite{angioni}
correlates favourably with our prediction $\hat{s}_{crit}\propto\delta W^{1/3}.$


\subsection{Neoclassical scaling}
It remains to explore the on-axis criterion introduced in Section III B.
Since, experimentally, sawtooth crashes are observed to occur when
$q_{0}\approx 0.75,$ \cite{chapman} we solve Eq. \eqref{eq:qnotnorm} for several
values of $\nu_{*,}$ ranging from $\nu_{*}=10^{-4}$ to $\nu_{*}=0.1$
and evaluate numerically the time at which $q(0,t)=0.75\,.$ These times are presented in Table \ref{tab:tab2}.
\begin{table}
\centering{}%
\begin{tabular}{|c|c||c|c|}
\hline 
$\nu_{*}$  & $\tau_{SAW}/\tau_{\eta}$  & $\nu_{*}$  & $\tau_{SAW}/\tau_{\eta}$\tabularnewline
\hline 
$0.1$  & $7.6\times10^{-3} $ & $0.01 $ & $2.3\times10^{-3}$ \tabularnewline
\hline 
$0.09 $ & $7.2\times10^{-3}$  & $ 0.005$  & $ 1.7\times10^{-3}$ \tabularnewline
\hline 
$0.075$  & $6.5\times10^{-3} $ & $ 0.0025$  & $ 1.3\times10^{-3} $\tabularnewline
\hline 
$0.05 $ & $5.3\times10^{-3} $  & $0.001 $ & $1.0\times10^{-3}$ \tabularnewline
\hline 
$0.03$  & $4.0\times10^{-3}$ & $ 0.0005$  & $ 0.90\times10^{-3}$ \tabularnewline
\hline 
$0.025$  & $ 3.6\times10^{-3}$ & $ 0.00025 $ & $0.84\times10^{-3}$ \tabularnewline
\hline 
$0.015$  & $2.8\times10^{-3}$ & $0.0001 $ & $0.80\times10^{-3}$ \tabularnewline
\hline 
\end{tabular}\caption{The ratio of $\tau_{SAW}$ to $\tau_{\eta}$ as a function of $\nu_{*e}$. These values are plotted in Fig. \ref{fig:fig4}.}
\label{tab:tab2} 
\end{table}
\begin{figure}
\includegraphics[scale=0.8]{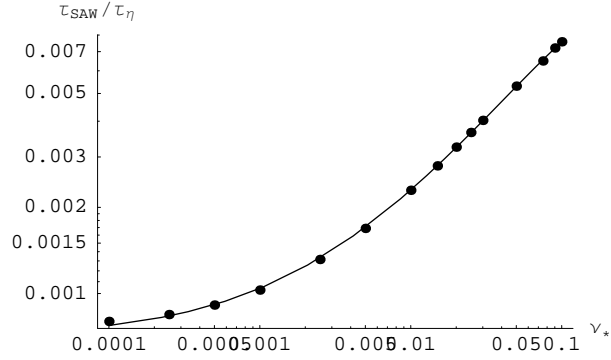} \caption{Data from Table \ref{tab:tab2}. A fit gives $\tau_{SAW}/\tau_{\eta}=7\times 10^{-4}+0.034\nu_{*}^{2/3}+0.004 \nu_*-0.08\nu_{*}^2$. }
\label{fig:fig4} 
\end{figure} Figure \ref{fig:fig4} shows the results of Table \ref{tab:tab2} graphically. 
A fit to the data is
\begin{equation}
\tau_{SAW}/\tau_{\eta}=7\times 10^{-4}+0.034\nu_{*}^{2/3}+0.004 \nu_*-0.08\nu_{*}^2.
\end{equation}
From this it is evident that when $0.001\lesssim\nu_{*}\lesssim 0.1,$ $\tau_{SAW}/\tau_{*}\sim\nu_{*}^{-1/3},$
so that
\begin{equation}
\tau_{SAW}\sim\tau_{*}\nu_{*}^{-1/3}\propto R_{0}^{8/3}N_{e}^{2/3}T_{e}^{1/6}\,\mbox{\mbox{sec}},\label{eq:myscaling}
\end{equation} where we do not distinguish between the lengths $R_0$ and $a,$ since JET and ITER share the same aspect ratio, $R_0/a=3.$
Equation (\ref{eq:myscaling}) shows a much weaker dependence of $\tau_{SAW}$ on $T_e$ than  for the resistive time scale, $\tau_{\eta}.$ 
For smaller values of $\nu_{*},$ $\tau_{SAW}$ scales as $\tau_*\nu_{*}^{-1}\sim\tau_{\eta},$ thus the $\nu_{*}$ dependence in $\tau_{SAW}$ disappears. This
is equivalent to saying that the fast evolution of $q$ on axis is a
transient phenomenon regularising the $q$-profile. For very
small values of $\nu_{*}$ such that $\delta_{*}/r_{1}\ll1,$ the effect
of $\nu_{*}$ on the global diffusion of the $q$-profile is negligible.
However numerically it is a much faster process than resistive diffusion, in agreement with Ref. \cite{parkmonticello}. 
From a preliminary analysis, we also find that the presence of the Bootstrap current terms in Ohm's law reduces the strength of the electron trapping effect, i.e. the development of localised axial structures near the axis. 
\begin{table}[hbtp]
 \centering{}%
\begin{tabular}{|c||c|}
\hline 
JET  & ITER\tabularnewline
\hline 
$a=1\,\mbox{m}$  & $a=3\,\mbox{m}$\tabularnewline
\hline 
$T_{e}=4\,\mbox{keV}$  & $T_{e}=25\,\mbox{keV}$\tabularnewline
\hline 
$\tau_{\eta}\sim400\,\mbox{sec}$  & $\tau_{\eta}\sim24\times10^{3}\,\mbox{sec}$\tabularnewline
\hline 
$\tau_{*}\sim0.86\,\mbox{sec}$  & $\tau_{*}\sim3\,\mbox{sec}$\tabularnewline
\hline 
$\nu_{*}\sim0.01\,$  & $\nu_{*}\sim6\times10^{-4}\,$\tabularnewline
\hline 
$\delta_{*}\sim4.6\,\mbox{cm}$  & $\delta_{*}\sim1.4\,\mbox{cm}$\tabularnewline
\hline 
\end{tabular}\caption{Resistive time $\tau_{\eta}$, fast diffusive time $\tau_{*}$, normalised
electron collision frequency $\nu_{*},$ and boundary layer $\delta_{*}$
for both JET and ITER. }
\label{tab:tab3}. 
\end{table}
In our case, when we consider the ratio $\tau_{\eta}/\tau_{*},$ we
see that $\tau_{\eta}^{JET}/\tau_{*}^{JET}\sim10^{2},$ and $\tau_{\eta}^{ITER}/\tau_{*}^{ITER}\sim10^{4}.$
If we take $R_{0}/a=3,$ and $N_{e}\sim10^{20}m^{-3}$ and compare
the two machines, we obtain the results in Table \ref{tab:tab3}. By using the results in Tables \ref{tab:tab2} and \ref{tab:tab3}, one obtains $\tau_{SAW}^{JET}\approx 1.69\, \mbox{sec},$ and  $\tau_{SAW}^{ITER}\approx 25\, \mbox{sec}.$ A sawtooth period of $1.7\,\mbox{sec.}$ is the longest observed in JET, while $40\,\mbox{sec.}$  is the value empirically allowed to avoid triggering Neoclassical Tearing Modes in ITER \cite{chapman2}.

\section{Discussion and Conclusions}

In this work we have discussed the role of the dissipative, $m=1,\,n=1,$ modes, in determining the sawtooth period in tokamaks, and explored the effect of neoclassical resistivity in the evolution of the plasma during the quiescent ramp phase of the sawtooth. 

In Ref. \cite{connor-hastie-zocco}, we calculated the critical value of the stability index $\Delta^{\prime}_{1,1}$ for crossing a linear stability threshold of the drift-tearing and dissipative kink modes, with gyrokinetic ions and semicollisional electrons. The stability thresholds derived in Ref. \cite{connor-hastie-zocco} depend sensitively on the magnitude of certain plasma parameters at the $q=1$ surface, such as the shear $\hat s_1(t),$ $\hat\beta(t)=0.5\beta_eL_s^2/L_n^2,$ and $\eta_e.$ We have therefore explored the resistive evolution of  $\hat s_1(t),$ and $\hat\beta(t)$. In addition, in order to address the possibility that the $m=1,\,n=1$ mode might actually be triggered by a core plasma instability near the magnetic axis, we have also monitored the evolution of $q_0(t).$

For the parameters defining the linear stability threshold of the  $m=1,\,n=1$ mode, [i.e.  $r_1(t),$ $\hat s_1(t), $ and $\hat \beta(t),$] we found negligible dependence on the collisionality parameter $\nu_*,$ but faster evolution [in agreement with Ref. \cite{parkmonticello}] than would occur with Spitzer resistivity. On this basis, one would expect a scaling of the sawtooth period from JET to ITER proportional to $T_e^{3/2}a^2.$ 

However, if the rapid downward evolution of $q_0(t)$ were to be responsible for triggering a sawtooth collapse, we find some sensitivity to the magnitude of $\nu_*,$ a weaker scaling of the sawtooth period with $T_e,$ a new scaling with the electron density $N_e$, and a different scaling with the machine size. These features may offer a means to distinguish the two different scalings using data from several machines.
A suggested scaling from JET to ITER, in this scenario, is $\tau_{SAW}\propto R_0^{8/3}N_e^{2/3}T_e^{1/6}.$ Notice that the weak temperature dependence we found in Eq. \eqref{eq:myscaling} mainly arises from the different dependencies of $\eta$ and $\nu_*$ on temperature. Such an axial criterion only requires a neoclassical post-crash evolution of the $q-$profile \cite{kelliher,lloyd}.
These results can be compared to the sawtooth $\tau_{SAW}\propto T^{3/2}R_0^2,$ period scaling suggested by Park and Monticello \cite{parkmonticello}, and also to the experimental data analysed by McGuire and Robinson \cite{mcguire} who found  the scaling $\tau_{SAW}\propto N_{e}^{3/7}T_{e}^{19/14}$ if resistive MHD equations govern the process, or $\tau_{SAW}\propto N_{e}^{3/5}T_{e}^{23/10},$ when diamagnetic effects were taken into account; these scalings were obtained with an empirical fit to data using appropriate dimensionless quantities.  The density dependence $\tau_{SAW}\propto N_e^{2/3}$ in Eq. \eqref{eq:myscaling} is similar to that found in Ref. \cite{mcguire}
 
While it is desirable to run simulations of the sawtooth
cycle that couple stability criteria and transport evolution of all
plasma profiles, in this work we contented ourselves with the analysis
of the post-crash, neoclassical $q$ evolution. In particular, the
simplified version of the resistivity we employed was helpful in identifying more directly the different
phases of the safety factor evolution during a sawtooth ramp. Finally,
we must stress that, while we calculated the exact relation between
critical shear and the ideal MHD potential energy $\hat s_{crit}^3\propto\delta W$, the
actual calculation of $\delta W$ and the study of the physical effects
that can change it are beyond the scope of this work.

\section*{Acknowledgements}

We are grateful to J B Taylor for several discussions that greatly improved our work.
This work was partly funded by the RCUK Energy Programme under grant
EP/I 501045 and the European Communities under the contract of Association
between Euratom and CCFE. A. Z. was supported by the Leverhume Trust
Network for Magnetised Plasma Turbulence, and a Culham Fusion Research
Fellowship. The views and opinions expressed herein do not necessarily
reflect those of the European Commission.


\end{document}